\documentstyle[pre,aps,multicol,eqsecnum,graphicx]{revtex}
\newcommand{\bleq}{\ifpreprintsty \else
\end{multicols}\vspace*{-3.5ex}{\tiny \noindent\begin{tabular}[t]{c|}
\parbox{0.493\hsize}{~} \\ \hline \end{tabular}} \fi}
\newcommand{\eleq}{\ifpreprintsty \else
{\tiny\hspace*{\fill}\begin{tabular}[t]{|c}\hline
\parbox{0.49\hsize}{~} \\
\end{tabular}}\vspace*{-2.5ex}\begin{multicols}{2} \fi}
\newcommand{\bcols}{\ifpreprintsty\else\begin{multicols}{2}\fi}
\newcommand{\ecols}{\ifpreprintsty\else\end{multicols}\fi}

\renewcommand{\vec}[1]{{\bf #1}} \renewcommand{\mathbf}{\bf\mathsf}
\let\oldhat=\hat \renewcommand{\hat}[1]{\oldhat{\bf #1}}

\newcommand{\ds}{\displaystyle}  \newcommand{\dst}{}

\begin{document}
\draft

\title{Pairing of Lyapunov Exponents for a Hard-Sphere Gas under Shear
in the Thermodynamic Limit}  \author{Debabrata Panja$^{*}$ and Ramses
van Zon$^{\dagger}$}  \maketitle
\begin{center}
{\it $^{*}$ Instituut Lorentz, Universiteit Leiden, \\ Postbus 9506,
2300 RA Leiden, The Netherlands.\\ $^{\dagger}$ Chemical Physics
Theory Group, Chemistry Department,  University of Toronto,  \\ 80
St.\ George St., Toronto, Ontario M5S 3H6, Canada. \/}
\end{center}
\begin{abstract}
\noindent
We consider a dilute gas of hard spheres under shear. We use one of
the predominant models to study this system, namely the SLLOD
equations of motion, with an iso-kinetic Gaussian thermostat in
between collisions, to get a stationary total peculiar kinetic energy.
Based on the previously obtained result that in the non-equilibrium
steady state and  in case the number of particles $N$ becomes large,
the coefficient of dynamical friction representing the iso-kinetic
Gaussian thermostat for the SLLOD dynamics fluctuates with
$1/\sqrt{N}$ fluctuations around a fixed value, we show on analytical
grounds that for a hard sphere gas at small shear rate and with a
large number of spheres, the conjugate pairing of the Lyapunov
exponents is expected to be violated at the fourth power of the
constant shear rate in the bulk.

\end{abstract}

\pacs{PACS Numbers: 05.20.-y, 05.45.-a, 05.60.Cd}

\bcols
\noindent
\section{Introduction}

Non-equilibrium molecular dynamics (NEMD) simulations of Navier-Stokes
equations have been used to study the shear viscosity properties of
fluids for a long time. To study the coefficient of shear viscosity in
Navier-Stokes equations, a carefully chosen periodic boundary
condition in NEMD simulations is enough to drive the system out of
equilibrium. Based on these ideas, in the beginning days of the
development of this subject, an algorithm was constructed from simple
Newtonian equations of motion using a periodic boundary conditions, the
so-called Lees-Edwards boundary conditions \cite{LE_jpc_72}. However,
it was soon realized that in the absence of an explicit dependence on
the shear field in this algorithm, one could not make an appropriate
connection with the Green-Kubo  relations, and therefore, it was
difficult to deal with from a theoretical point of view. As a remedy,
some other algorithm with an explicit dependence on the shear field
was called  for, and the DOLLS and the SLLOD algorithms were born.

The main idea behind the DOLLS and the SLLOD algorithm is an explicit
dependence on the shear  field, $\gamma$. The DOLLS algorithm was
implemented first \cite{HEHLAM_pra_80}. The SLLOD equations of motion
were proposed soon after \cite{EM_84}, and are now preferred because
they are equivalent to the boundary driven method
\cite{EM_book_academic}.  Both algorithms have to be supplemented by a
thermostat, which continuously removes the energy generated due to the
work done on the system by the shear field such that an
non-equilibrium steady state (NESS), homogeneous in space, can be
reached.

In this paper, we will look at the SLLOD equations of motion  for a
gas of hard spheres from the point of view of dynamical systems. The
Lyapunov exponents of a system of particles, obeying the SLLOD
equations of motion and mutually interacting by means of WCA potential
with an iso-kinetic Gaussian thermostat, was first studied by Morriss
\cite{Morriss_pla_89}. The study showed that the shear viscosity can
be obtained from the sum of all the Lyapunov exponents. The simulation
results in Ref.~\cite{Morriss_pla_89} also indicated that once the
Lyapunov exponents are arranged in ascending order of magnitude, the
sum of the largest and the smallest, the sum of the second largest and
the second smallest and so on, were the  same. The phenomenon of such
pairing of the Lyapunov exponents is known as the Conjugate Pairing
Rule, or the CPR. Since it is in general difficult to calculate all
the individual Lyapunov exponents of a system, an extensive
theoretical study  soon ensued to understand the CPR for the Lyapunov
exponents of systems obeying the SLLOD equations of motion.
Evans and co-workers investigated this point \cite{ECM_pra_90} to
conclude that the Lyapunov exponents pair exactly for general
inter-particle potentials and {\it all} $\gamma$. In a follow-up work,
Sarman {\it et al.\/} \cite{SEM_pra_92} carried out simulation studies
in support of Ref. \cite{ECM_pra_90}.

In the next few years, the connection between the dynamical systems
theory and statistical mechanics saw a surge of interest. Some
situations were found, where  it was possible to prove that the CPR is
satisfied exactly
\cite{DM_pre_96,WL_cmp_98,Ruelle_jsp_99,Panja_preprint}. The status of
the CPR for the SLLOD and the DOLLS dynamics was revisited.  For a
system of particles obeying the SLLOD and the DOLLS dynamics with a
WCA inter-particle potential and arbitrary $\gamma$, CPR was reported
to be violated on the basis of simulation
results\cite{ISE_physica_97,SEI_chaos_98}, but recently it was shown
that this claim is based on an erroneous analysis of the Lyapunov
spectrum\cite{morrisscomment}. However, for these two systems, no
attempt of a {\it theoretical\/} understanding about the nature of an
approximate CPR has been carried out  so far. In this paper, we
address and attempt to clarify these issues. We find that for a dilute
gas of hard spheres obeying the SLLOD dynamics, where the masses and
the radii of the spheres are not necessarily the same,  and the total
peculiar kinetic energy is kept constant by applying the iso-kinetic
Gaussian thermostat in between collisions, the CPR is violated at the
most at $O(\gamma^{4})$, for small $\gamma$, in the thermodynamic
limit.  \cite{Noteonthermodynamiclimit} Our analysis is based on the
key idea that the coefficient of friction representing the iso-kinetic
Gaussian thermostat for a  dilute gas of particles mutually
interacting by means of a short-ranged potential and obeying the SLLOD
dynamics with a small shear rate, in the NESS, reaches a fixed value
in the thermodynamic limit, with $1/\sqrt{N}$ fluctuations, where $N$
is the number of particles.  \cite{Zon_pre_99}.

The structure of the paper is the following: in Sec.~II, we describe
equations of motion for the SLLOD dynamics, define the Lyapunov
exponents, and discuss the  sufficient conditions for an exact CPR. In
Sec.~III, we demonstrate how the coefficient of friction representing
the iso-kinetic Gaussian thermostat is expected to reach a fixed value
at the thermodynamic limit, with $1/\sqrt{N}$ fluctuations. In
Sec.~IV, we present the explicit calculations and discuss the status
of an approximate CPR. To make the calculations In Secs.~II-IV simple,
we assume that each of the gas particles has a unit mass. Finally, we
end this paper with discussions on possible generalizations, including
the generalization to the case when the masses of the gas particles
are arbitrary, in Sec.~V.

\section{The SLLOD Equations of Motion from a Viewpoint
of Dynamical Systems}

\subsection{Equations of Motion}

The SLLOD equations of motion describe the dynamics of a collection of
$N$ particles constituting a fluid with a macroscopic velocity field
$\vec u=\gamma y \hat x$ (i.e., the gradient of the $x$-component of
the macroscopic fluid velocity $\vec u$ in the $y$-direction is
$\gamma$). For simplicity, each gas particle is assumed to have a unit
mass. The specific form of the SLLOD equations of motion for the
$i$-th particle,  in terms of its position $\vec{r}_{i}$ and peculiar
momentum $\vec{p}_{i}$, is given by
\begin{eqnarray}
    \dot{\vec{r}_{i}}= \vec{p}_{i}\,+\,\gamma y_{i}\hat{x}\,, \quad
	\dot{\vec{p}_{i}}= \vec{F}_{i}-\gamma
	p_{iy}\hat{x}-\alpha\vec{p}_{i}\,.
\label{e2.1}
\end{eqnarray}
where $\vec{F}_{i}$ is the force on the $i$-th particle due to the
other particles in the system. The peculiar velocity of a particle is
defined as its velocity with respect to the velocity of the flow at
its location and the peculiar momentum of a particle is the product of
its mass and its peculiar velocity. The value of $\alpha$, the
coefficient of friction representing the iso-kinetic Gaussian
thermostat in Eq.~(\ref{e2.1}), is chosen such that the total peculiar
kinetic energy of the system, $\sum_{i}p^{2}_{i}/2$, is a constant of
motion, i.e.,
\begin{eqnarray}
   \alpha= \frac{\sum_{i=1}^{N}\,(\vec{F}_{i}\cdot\vec{p}_{i}-\gamma
	p_{ix}p_{iy})}{\sum_{i=1}^{N}\,p^{2}_{i}}\,.
\label{e2.2}
\end{eqnarray}
The SLLOD equations of motion, without the dissipative term
$-\alpha\vec{p}_i$, cannot be derived from a Hamiltonian (unlike the
DOLLS equations of motion).

We will use the equations of motion exclusively in terms of the
particles' positions $\vec{r}_{i}$ and laboratory velocity
$\vec{v}_{i}$. This introduces the change of variable from
$\vec{p}_{i}$ to $\vec{v}_{i}=\vec{p}_{i}+\gamma y_{i}\hat{x}$ in
Eq.~(\ref{e2.1}), which can then be written as
\begin{eqnarray}
   \dot{\vec{r}_{i}}= \vec{v}_{i}\,,  \quad	 \dot{\vec{v}_{i}}=
\vec{F}_{i}\,+\,\alpha\gamma y_{i}\hat{x}-\alpha\vec{v}_{i}\,.
\label{e2.3}
\end{eqnarray}

In the present context, the gas particles are hard spheres of
arbitrary radii. This reduces the dynamics of the gas particles to an
alternating sequence of flight segments and instantaneous binary
collisions. During a flight, the dynamics of the gas particles is
therefore described by  Eqs.~(\ref{e2.2}-\ref{e2.3}) with
$\vec{F}_{i}=0$.  As far as the dynamics in collisions is concerned,
it is possible to derive the limiting behavior of the iso-kinetic
Gaussian thermostat as $\vec F_i\rightarrow\infty$\cite{Petravic}, but
this leads to rather complicated collision rules. For the purpose of
simplicity, in this paper, we choose to apply the iso-kinetic
thermostat only {\em in between collisions.} Thus, at an instantaneous
collision between the $i$-th and the $j$-th sphere
($i,j=1,2,..,N;\,i\neq j$), the post-collisional positions and
laboratory momenta ($+$ subscripts) are related to their
pre-collisional values ($-$ subscripts) by
\begin{eqnarray}
   \vec{r}_{i+}&=& \vec{r}_{i-}, \hspace{1cm} \vec{r}_{j+}=
	\vec{r}_{j-}\,,
\label{e2.4}
\\ \vec{v}_{i+}&=& \vec{v}_{i-}-\{(\vec{v}_{i-}-\vec{v}_{j-})\cdot
   \hat {n}_{ij}\}\,\hat{n}_{ij}
\label{e2.5}
\end{eqnarray}
and
\begin{eqnarray}
   \vec{v}_{j+}&=& \vec{v}_{j-}\,+\,\{(\vec{v}_{i-}-\vec{v}_{j-})\cdot
	\hat{n}_{ij}\}\,\hat{n}_{ij}\,,
\label{e2.6}
\end{eqnarray}
while the positions and the velocities of the rest of the spheres
remain unchanged. Here, $\hat{n}_{ij}$ is the unit vector along the
line joining the center of the $i$-th sphere to the $j$-th sphere at
the instant of collision. Note that although in any particular
collision, the peculiar kinetic energy changes over a collision, these
changes are random, both in magnitude and sign, due to the randomness
of the collision parameters, and hence it is quite likely that the
system would reach a steady state, where the average change of
peculiar kinetic energy would be zero.

To study the SLLOD dynamics as a dynamical system in three dimensions
(the dimensionality does not affect our analysis), we form the
$3N$-dimensional vectors
$\vec{R}=(\vec{r}_{1},\vec{r}_{2},...,\vec{r}_{N})$,
$\vec{V}=(\vec{v}_{1},\vec{v}_{2},...,\vec{v}_{N})$ and
$\hat{N}_{ij}$, whose $l$-th entry is given by $\hat{N}^{l}_{ij}=
(\delta_{l,i}-\delta_{l,j})\,\hat{n}_{ij}/\sqrt{2}\,\,\,\,(l=1,2,..,N)$.
Using these new variables, we write the SLLOD equations of motion in
the $(\vec{r}_{i},\vec{v}_{i})$ coordinates of $N$ hard spheres during
a flight, Eq.~(\ref{e2.3}), in a compact form
\begin{eqnarray}
   \dot{\vec{R}}= \vec{V}\,, \quad \dot{\vec{V}}=\,\alpha\gamma
	{\mathbf C} \vec{R}-\alpha\vec{V}\,.
\label{e2.7}
\end{eqnarray}
Here, ${\mathbf C}$ is a $3N\!\times\!3N$ matrix with $N\!\times\!N$
entries, each of which is a $3\!\times\!3$ matrix. In terms of the
entry index $(l,m)$, in the $xyz$-basis, ${\mathbf C}_{lm}={\mathbf
c}\,\delta_{lm}$ ($l,m=1,2,..,N$) and
\begin{eqnarray}
   {\mathbf c}= \hat x\hat y = \left[\begin{array}{ccc} 0&1&0\\
	 0&0&0\\ 0&0&0 \end{array}\right]\,.
\label{e2.8}
\end{eqnarray}
At a collision between the $i$-th and the $j$-th sphere, the equations
of motion are given by \cite{Panja_preprint}
\begin{eqnarray}
   \vec{R}_{+}= \vec{R}_{-}\,, \quad \vec{V}_{+}=
   \vec{V}_{-}-2\,(\vec{V}_{-}\cdot\hat{N}_{ij})\,\hat{N}_{ij}\,.
\label{e2.9}
\end{eqnarray}
In our analysis hereafter, except for Sec.~III, we will use only
Eqs.~(\ref{e2.7}-\ref{e2.9}) to describe the dynamics.

At this point, we introduce the following notations. A phase space
point can be denoted as $\vec\Gamma=(\vec R,\vec V)$. A linear
transformation on phase space can be given as a $6N\times 6N$
matrix. Any such matrix $\mathbf P$ can be split in terms of four
$3N\times 3N$ sub-blocks, for which we use the notation  ${\mathbf
P}^{[1]}$, ${\mathbf P}^{[2]}$, ${\mathbf P}^{[3]}$ and  ${\mathbf
P}^{[4]}$,  such that
\begin{equation}
   {\mathbf P} = \left[\begin{array}{cc} {\mathbf P}^{[1]} & {\mathbf
	P}^{[2]} \\ {\mathbf P}^{[3]} & {\mathbf P}^{[4]}
	\end{array}\right]\,.
\label{e2.10}
\end{equation}
Each  sub-block ${\mathbf P}^{[i]}$ ($i=1,\ldots,4$) itself can be
divided in $3\times 3$  sub-blocks again, where each sub-block can be
identified by two indices $l$ and $m$, $l$ along the horizontal and
$m$ along the vertical direction ($l,m=1,2,..,N$). Such a sub-block is
denoted by ${\mathbf P}^{[i]}_{lm}$.

\subsection{Lyapunov Exponents for Hard-sphere Systems}

To calculate the Lyapunov exponents for hard-sphere systems, let us
say that the system starts at time $t_{0}$ at a phase-space location
$\vec{\Gamma}(t_{0})\equiv(\vec{R}(t_{0}),\vec{V}(t_{0}))$. Under time
evolution, $\vec{\Gamma}(t)$ follows a trajectory in the
$6N$-dimensional phase space, which we call the ``reference
trajectory''. The set of $N$ hard spheres would suffer a sequence of
binary collisions on this trajectory. We also consider an
infinitesimally displaced trajectory in the phase space, which starts
at the same time $t_{0}$, but at $\vec{\Gamma}'(t_{0})
=\vec{\Gamma}(t_{0})+\vec{\delta\Gamma}(t_{0})$. Under time evolution,
$\vec{\Gamma}'(t)$ follows another trajectory, always staying
infinitesimally close to the reference trajectory. This trajectory we
call the ``adjacent trajectory''. We also assume that the set of $N$
hard spheres on the reference and the adjacent trajectories suffer the
{\it same\/} sequence of binary collisions. We denote the time
evolution of the infinitesimal $6N$-dimensional  {\it tangent vector}
$\vec{\delta\Gamma}(t)$ over time $(t-t_{0})$ by the $6N\!\times\!6N$
matrix ${\mathbf L}(t-t_{0})$, i.e.,
\begin{eqnarray} 
   \vec{\delta\Gamma}(t)= {\mathbf
	L}(t-t_{0})\,\vec{\delta\Gamma}(t_{0})\,.
\label{e2.11}
\end{eqnarray}
The Lyapunov exponents are the possible exponential growth rates in
time of $|{\mathbf L}\hat \Gamma|$ for different directions of unit
vectors $\hat \Gamma$. We have to define the norm in an appropriate
way. Making the time it takes for a sphere with a typical velocity
$v_0$ to cross the distance of a typical radius of a sphere $a_0$ our
unit of time (i.e., $a_0/v_0$ is set to $1$) solves the problem that
the components of $\hat\Gamma$ have different dimensions. For the
inner product between two tangent vectors
$\vec{\delta\Gamma}^{(1)}=(\vec{\delta R}^{(1)},\vec{\delta V}^{(1)})$
and $\vec{\delta\Gamma}^{(2)}=(\vec{\delta R}^{(2)},\vec{\delta
V}^{(2)})$  we use
\[
\langle\vec{\delta \Gamma}^{(1)}|\vec{\delta\Gamma}^{(2)} \rangle=
\sum_{i=1}^N (\vec{\delta r}^{(1)}_i\cdot\vec{\delta r}^{(2)}_i+
\vec{\delta v}^{(1)}_i\cdot\vec{\delta v}^{(2)}_i).
\]
The norm is now defined as $|\vec{\delta\Gamma}|=\sqrt
{\langle\vec{\delta \Gamma}|\vec{\delta\Gamma}\rangle}$.

The Lyapunov exponents are the logarithms of the eigenvalues of the
matrix ${\mathbf\Lambda}$, defined by
\begin{eqnarray} 
   {\mathbf \Lambda}= \lim_{t\rightarrow\infty}\left[
	{\mathbf\tilde{L}}(t-t_{0})\right]^{1/[2(t-t_{0})]}\,,
\label{e2.12}
\end{eqnarray} 
where, ${\mathbf \tilde{L}}(t-t_{0}) =
[{\mathbf{L}}(t-t_{0})]^{\mbox{\scriptsize T}}\,
{\mathbf{L}}(t-t_{0})$.  The corresponding directions for the
exponential expansion and contraction of the phase space
$(\vec{R},\vec{V})$  are obtained from the eigenvectors of ${\mathbf
\tilde{L}}(t-t_{0})$.

The dynamics of $\vec{\delta\Gamma}(t)$ in Eq.~(\ref{e2.11}), for a
gas of hard spheres, can be decomposed into an alternating sequence of
flights and instantaneous binary collisions.  We denote the
transformation of $\vec{\delta\Gamma}(t)$ over a flight segment
between $t$ and $t+\Delta t$ by ${\mathbf H}(\Delta t)$ such that
\begin{eqnarray}
   \vec{\delta\Gamma}(t+\Delta t)= {\mathbf H}(\Delta
	t)\,\vec{\delta\Gamma}(t)
	\hspace{0.4cm}\mbox{with}\hspace{0.4cm}{\mathbf H}(0)={\mathbf
	I}\,.
\label{e2.13}
\end{eqnarray}
Explicitly, ${\mathbf H}(\Delta t)$ is obtained from
\begin{eqnarray}
   \dot{\vec{\delta\Gamma}}(t) = {\mathbf T}(t)\,\vec{\delta\Gamma}(t)
\label{e2.14}
\end{eqnarray}
as
\begin{eqnarray}
{\mathbf H}(\Delta t)=
\exp_{\tiny\mbox{\,T}}\,\left[\int_{t}^{t+\Delta t}dt'\, {\mathbf
T}(t')\right]\,,
\label{e2.15}
\end{eqnarray}
where the subscript T indicates time ordering. Notice that ${\mathbf
H}(\Delta t)$ in a general system will depend on  time $t$ as well,
but we have suppressed that in our notation.  If we now denote the
transformation of $\vec{\delta\Gamma}(t)$ over an instantaneous binary
collision (say, between the $i$-th and the $j$-th sphere) by the
matrix  ${\mathbf M}_{ij}$,  we can express the matrix ${\mathbf
L}(t-t_{0})$ in terms of the ${\mathbf H}$ and  ${\mathbf M}_{ij}$
matrices in the following way: if the dynamics involves  flight
segments separated by $s$ instantaneous binary collisions at $t_{1}$,
$t_{2}\ldots t_{s}$ such that $t_{0}<t_{1}<t_{2}<\ldots<t_{s}<t$, then
\begin{eqnarray}
   {\mathbf L}(t-t_{0})= {\mathbf H}(\Delta t_{s})\,{\mathbf
	M}_{i_sj_s}\, {\mathbf H}(\Delta t_{s-1}) \ldots\nonumber \\&&
	\hspace{-2.7cm} \ldots {\mathbf M}_{i_1j_1}\,{\mathbf
	H}(\Delta t_{0})\,.
\label{e2.16}
\end{eqnarray}
Here, $\Delta t_i=t_{i+1}-t_i$ for $i=1,2,\ldots,(s-1)$ and $\Delta
t_s=t-t_s$.

\subsection{The Sufficient Conditions for an Exact CPR}

If the CPR is exactly satisfied for  a dynamical system, the sum of
the conjugate pairs of the Lyapunov exponents is  some constant $c$,
i.e., if $\lambda_{i}$ is a Lyapunov exponent of this system, then
$c-\lambda_i$ is also a Lyapunov exponent.  The proof of a possible
conjugate pairing rule will follow from the properties of the matrix
${\mathbf L}(t-t_{0})$. However, to understand the interplay between
the properties of the matrix ${\mathbf L}(t-t_{0})$ and an exact CPR
in full generality, below we first look at the property of ${\mathbf
L}(t-t_{0})$ that has been used in various cases to prove CPR.

\begin{enumerate}
\item[(a)]  If the matrix ${\mathbf L}(t-t_0)$ is symplectic, i.e.,
${\mathbf L}(t-t_0)$ satisfies the symplectic condition
\[
[{\mathbf L} (t-t_0)]^{\mbox{\scriptsize T}}{\mathbf J}{\mathbf
L}(t-t_0)= {\mathbf J},
\] 
with ${\mathbf J}$ as the usual symplectic matrix, then
\begin{eqnarray}
   {\mathbf\tilde{L}}(t-t_{0})\, {\mathbf
	J}\,{\mathbf\tilde{L}}(t-t_{0})= {\mathbf J}\,.
\label{e2.17}
\end{eqnarray}
Eq.~(\ref{e2.17})  can be used to show that
\begin{eqnarray}
\mbox{Det}\,[\,{\mathbf\tilde{L}}(t-t_{0})\,-\,\tilde{L}\,{\mathbf
	I}\,] =0\nonumber
\end{eqnarray}
(with $\mathbf I$ the identity matrix) implies
\begin{eqnarray}
\mbox{Det}\,[\,{\mathbf\tilde{L}}(t-t_{0})
-\frac{1}{\tilde{L}}\,{\mathbf I}\,] =0.
\label{e2.18}
\end{eqnarray}
This means that if $\tilde{L}$ is an eigenvalue of
${\mathbf\tilde{L}}(t-t_{0})$ then so is $\tilde{L}^{-1}$.  Since the
Lyapunov exponents are the logarithms of the eigenvalues of the matrix
${\mathbf\Lambda}$, defined in Eq.~(\ref{e2.12}), it is easy to see
that $c=0$. All Hamiltonian systems fall in this class.
\item[(b)] In the existing literature
\cite{DM_pre_96,WL_cmp_98,Ruelle_jsp_99,Panja_preprint}, the concept
of the symplectic condition defined above, has been generalized to the
so-called ``$\mu$-symplectic condition'' and applied to thermostatted
systems where an iso-kinetic Gaussian thermostat keeps the total {\it
laboratory\/} kinetic energy constant and the external force on the
constituent particles of the system is dependent only on the positions
of the particles. For these systems, in an appropriate reduced phase
space characterized by all the non-zero Lyapunov exponents, the matrix
${\mathbf L}(t-t_{0})$ satisfies this $\mu$-symplectic condition,
which means that  there exists a {\it time-dependent positive
scalar\/} quantity $\mu$, such that $[{\mathbf
L}(t-t_{0})]^{\mbox{\scriptsize T}}{\mathbf J}{\mathbf L}(t-t_{0})=\mu
{\mathbf J}$.  This implies that
\begin{eqnarray}
    {\mathbf\tilde{L}}(t-t_{0})\,{\mathbf J}
    {\mathbf\tilde{L}}(t-t_{0}) = \mu^{2}{\mathbf J}\,;
\label{e2.19}
\end{eqnarray}
which can be used to derive that if
\begin{eqnarray}
	\mbox{Det}[{\mathbf\tilde{L}}(t-t_{0})-\tilde{L} {\mathbf I}]
	=0\nonumber
\end{eqnarray}
then also
\begin{eqnarray}
   \mbox{Det}[{\mathbf\tilde{L}}(t-t_{0})-\frac{\mu^{2}}{\tilde{L}}
   \,{\mathbf I}\,] =0\,,
\label{e2.20}
\end{eqnarray}
for an eigenvalue $\tilde{L}$ of the matrix
${\mathbf\tilde{L}}(t-t_{0})$. That means that if $\tilde{L}$ is an
eigenvalue of ${\mathbf\tilde{L}}(t-t_{0})$ then so is
$\mu^{2}\tilde{L}^{-1}$. In that case, one finds from
Eq.~(\ref{e2.12}) that
$c=\!\dst{\lim_{t\rightarrow\infty}}(\ln\mu)/(t-t_{0})$.  If the
system is ergodic, then this long time average for $c$ can be equated
to a NESS average. Notice that condition (a) is obtained as special
case of condition (b), namely when $\mu=1$.
\end{enumerate}
\noindent
Returning momentarily to the SLLOD dynamics, we observe that the
formalism developed in
Refs.~\cite{DM_pre_96,WL_cmp_98,Ruelle_jsp_99,Panja_preprint} fails
here. The primary reason is associated with the fact that the total
{\it peculiar\/} kinetic energy is held constant for the SLLOD
dynamics, as opposed to the total {\it laboratory\/} kinetic energy in
Refs.~\cite{DM_pre_96,WL_cmp_98,Ruelle_jsp_99,Panja_preprint}. One
however needs to interpret the statement regarding the connection
between the violation of condition (b) and the non-exactness of the
CPR with care. By virtue of the fact that condition (b) above is a
sufficient condition for the CPR to hold exactly, the violation of an
exact CPR {\it cannot be guaranteed\/} if condition (b) is not
satisfied.

Guided by this observation,  the interplay between the properties of
the matrix ${\mathbf L}(t-t_{0})$ and an exact CPR for a dynamical
system can be generalized further than what is presented in (b). If
there exists {\it any constant non-singular\/} matrix ${\mathbf K}$
satisfying ${\mathbf K}^{2}\propto{\mathbf I}$ and the following
condition
\begin{eqnarray}
    [{\mathbf L}(t-t_{0})]^{\mbox{\scriptsize T}}\,{\mathbf K}\,
	{\mathbf L}(t-t_{0}) =\mu{\mathbf K}
\label{e2.21}
\end{eqnarray}
is satisfied with a time-dependent scalar quantity $\mu$, then
Eq.~(\ref{e2.20}) can be shown to hold for an eigenvalue $\tilde{L}$
of the matrix ${\mathbf\tilde{L}}(t-t_{0})$, implying that the CPR is
exactly satisfied for such a dynamical system \cite{sufficient}.  In
analogy with the nomenclature presented in (b), we call
Eq.~(\ref{e2.21}) a ``generalized $\mu$-symplectic condition''  with
matrix ${\mathbf K}$.  We emphasize  that the necessary condition for
an exact CPR to hold for a dynamical system is not known.

In view of Eq.~(\ref{e2.21}), thus, one should look for such a matrix
${\mathbf K}$ to prove an exact CPR. Instead, we look at it from a
different angle, namely that we would like to understand how the SLLOD
dynamics of $N$ hard spheres with an iso-kinetic Gaussian thermostat
deviates from an exact CPR.

\section{Behavior of $\alpha$ in the Thermodynamic Limit}

Our procedure to study the deviations from CPR begins with the
following observation:   in the thermodynamic limit, for the SLLOD
dynamics with short-range inter-particle potentials  at low density of
spheres and at small $\gamma$, the behavior of $\alpha$ simplifies to
a great extent.  After some transient time the system reaches the
NESS, and the coefficient of friction $\alpha$ fluctuates with
$1/\sqrt{N}$ fluctuations around a fixed value $\alpha_{0}$
\cite{Zon_pre_99}. For not too large fluctuations, the distribution
function for $\alpha$ can also be shown to be approximately
Gaussian. Thus, to calculate the Lyapunov exponents for large $N$ at
low density of spheres and at small $\gamma$, {\it to which we confine
ourselves henceforth}, $\alpha$ can be replaced by $\alpha_{0}$ in
Eq.~(\ref{e2.7}). We will now briefly present the gist of the
derivation in Ref.~\cite{Zon_pre_99}, applied to hard spheres.

For a hard sphere system, the force term in  Eqs.~(\ref{e2.1}) and
(\ref{e2.2}) is zero during a flight. Thus, for a flight, we have
\begin{eqnarray}
   \alpha= -\,\gamma \left[ \sum_{i=1}^N \,p_i^2\right]^{-1}
	\sum_{i=1}^N\,p_{ix}p_{iy}\,.
\label{e3.1}
\end{eqnarray}
Introducing a second thermostat variable,
\begin{eqnarray}
   \beta= \gamma^2 \left[\sum_{i=1}^N\,p_i^2\right]^{-1}
	\sum_{i=1}^N\,p_{iy}^2\,,
\label{e3.2}
\end{eqnarray}
a closed set of equation follows from Eq.~(\ref{e2.1}):
\begin{eqnarray}
   \dot\alpha &=& -\,2\alpha^2\,+\,\beta  \nonumber \\ \dot\beta&=&
-\,2\alpha\beta\,.
\label{e3.3}
\end{eqnarray}
These equations are valid during the flights, i.e., the intervals
between collisions.

We treat collisions by looking at their net effect, i.e., how the
velocities and positions of the particles $i$ and $j$ involved in the
collision, are changed from their  pre-collisional values $\vec
p_{i-}$ and $\vec p_{j-}$ to their post-collisional values
$\vec{p}_{i+}$ and $\vec{p}_{j+}$. These are the only two velocities
to change, and because $\sum_{i=1}^N p_i^2$ is of order  $N$, the
changes in $\alpha$ and $\beta$ are, according to Eqs.~(\ref{e3.2})
and (\ref{e3.3}), of order $N^{-1}$.  The number of collisions in the
whole system is an extensive quantity as well, so there are $O(N)$ of
these $O(1/N)$ changes in a unit of time. The averages of the small
changes are not zero, so there is a net effect of $O(1)$ per unit time
to the time derivatives of $\alpha$ and $\beta$, which we will  denote
by $a$ respectively, $b$:
\begin{eqnarray}
   \dot\alpha\!&=&\!-\,2\alpha^2\,+\,\beta\,+\,a \nonumber \\
\dot\beta\!&=&\!-\,2\alpha\beta\,+\,b\,.
\label{e3.4}
\end{eqnarray}
This set of equations has a fixed point $(\alpha_0,\beta_0)$, which is
stable if $\alpha_0>0$, so the system reaches this fixed point after
some time. On top of this dynamics, there are fluctuations. Assuming
that the  collisions are independent, the central limit theorem
applies, and the fluctuations are $O(1/\sqrt{N})$. For more detailed
analysis, we refer to Ref.  \cite{Zon_pre_99}, where the independence
of the changes in $\alpha$ and $\beta$ is linked to the assumption of
molecular chaos.

Finally, we note that to maintain a stationary total peculiar kinetic
energy in a system with a constant $\alpha_0$ thermostat, this
constant has to be chosen differently for varying $\gamma$. For
$\gamma=0$, i.e., in equilibrium, $\alpha_0$ can be set to zero, and
kinetic energy is determined by the initial conditions. Near
equilibrium, i.e., in the linear response regime, the right hand side
of Eq.~(\ref{e3.1}) should scale as $\gamma^2$,
i.e. $\alpha_0\propto\gamma^2$. Obviously, there are higher order
corrections to this behavior which play a role for larger values of
$\gamma$. If, for $\gamma\neq0$, the initial condition is such that
the total peculiar kinetic energy is not equal to the stationary
value, that value will be approached in time.

\section{Status of an Approximate CPR in the Thermodynamic Limit}

Based on the discussion in the last paragraph of Sec.~II and using the
results in  Sec.~III, we will explore the possibility of an
approximate CPR for the the SLLOD dynamics of hard spheres in the
thermodynamic limit, at small $\gamma$ and at low density in this
section. We will first obtain the desired results for a constant
coefficient of friction $\alpha_{0}$ in the equations of motion
(\ref{e2.7}) and then we will discuss the validity of an approximate
CPR when the system is under an iso-kinetic Gaussian thermostat. To
this end, our starting aim is to study the generalized
$\mu$-symplectic properties of the matrix ${\mathbf L}(t-t_{0})$ for
the dynamics described by Eq.~(\ref{e2.7}) during a flight and
Eq.~(\ref{e2.9}) during an instantaneous binary collision between the
$i$-th and the $j$-th sphere.

However, as the matrix ${\mathbf L}(t-t_{0})$ is constructed from the
${\mathbf H}$ and the ${\mathbf M}$ matrices, we will have to study
the generalized  $\mu$-symplectic properties of the ${\mathbf H}$ and
the ${\mathbf M}$ matrices separately.

\subsection{Generalized $\mu$-symplecticity Property of ${\mathbf
H}(\Delta t)$}

The matrix ${\mathbf T}$ describing the  dynamics of
$\vec{\delta\Gamma}$  during flights is found from Eq.~(\ref{e2.7}) to
be
\begin{eqnarray}
   {\mathbf T}\equiv  \left[\begin{array}{cc} {\mathbf 0}&{{\mathbf
	I}}\\ {\alpha_{0}\gamma{\mathbf C}\,\,\,}
	&{-\,\alpha_{0}{\mathbf I}} \end{array}\right]\,.
\label{e4.1}
\end{eqnarray}
{}From Eq.~(\ref{e4.1}), it is straightforward to obtain
\begin{eqnarray}
   {\mathbf H}(\Delta t)= \exp\,[{\mathbf T}\Delta t]\,,
\label{e4.2}
\end{eqnarray} 
where the $3\times3$ sub-blocks of ${\mathbf H}$ are given by
${\mathbf H}^{[k]}_{lm}(\Delta t)= {\mathbf h}^{(k)}(\Delta
t)\,\delta_{l,m}$, where
\begin{eqnarray}
   {\mathbf h}^{(1)}(\Delta t)&=&  {\mathbf I} + \left[\,\gamma\Delta
	t-{ \frac{\gamma\,[1-\exp(-\,\alpha_{0}\Delta t)]}
	{\alpha_{0}}}\right]{\mathbf c}\,,  \nonumber  \\  {\mathbf
	h}^{(2)}(\Delta t)&=& \frac{1-e^{-\,\alpha_{0}\Delta
	t}}{\alpha_{0}} {\mathbf I} \nonumber\\&& +
	\frac{\gamma}{\alpha_0^2}\left[ \alpha_0\Delta t
	(1+e^{-\alpha_0\Delta t})-2+2e^{-\alpha_0\Delta t} \right]
	{\mathbf c} \,,  \nonumber  \\  {\mathbf h}^{(3)}(\Delta t)&=&
	\gamma\,[1-e^{-\alpha_{0}\Delta t}]\,{\mathbf
	c}\,\quad\mbox{and} \nonumber  \\  {\mathbf h}^{(4)}(\Delta
	t)&=& e^{-\alpha_{0}\Delta t}\!  \left\{\!{\mathbf I}
	-\gamma\left[\Delta
	t+{\frac{1}{\alpha_{0}}[1-e^{\alpha_{0}\Delta t}]}\right]
	{\mathbf c}\right\}\!.
\label{e4.3}
\end{eqnarray} 

Due to the complicated form of ${\mathbf H}(\Delta t)$, it is easier
to study its $\mu$-symplecticity properties in terms of the matrix
${\mathbf T}$.  This involves the task of finding a possible matrix
${\mathbf K}$ satisfying the condition
\begin{eqnarray}
   {\mathbf T}^{\mbox{\scriptsize T}}{\mathbf K}\,+\,{\mathbf
K}{\mathbf T}= \beta\,{\mathbf K}
\label{e4.4}
\end{eqnarray}
such that ${\mathbf K}^{2}\propto{\mathbf I}$. If such a  matrix
$\mathbf K$ exists, then  ${\mathbf H}(\Delta t)$ is generalized $\mu$
symplectic with that matrix, and
\begin{eqnarray}
    \mu= \exp\left[\int_{t}^{t+\Delta t}\beta\,dt'\right].
\label{e4.5}
\end{eqnarray}
Since ${\mathbf K}$ and ${\mathbf T}$ are constant matrices in the
present context, $\beta$ is also a constant.  Eq.~(\ref{e4.4})  can be
treated as a simple eigenvalue  equation to solve for the eigenvalue
$\beta$ and the eigenvector ${\mathbf K}$. We find that there exists a
matrix ${\mathbf G}$ satisfying
\begin{eqnarray}
   {\mathbf T}^{\mbox{\scriptsize T}} {\mathbf G}\,+\,{\mathbf
G}{\mathbf T} =-\alpha_{0}\,{\mathbf G}
\label{e4.6}
\end{eqnarray}
and the $6N\!\times\!6N$ matrix ${\mathbf G}$  in terms of its
$3\times3$ sub-blocks,  is given by ${\mathbf G}^{[1]}={\mathbf
G}^{[4]}={\mathbf 0}$ and ${\mathbf G}^{[2]}_{lm}=-{\mathbf
G}^{[3]}_{lm}={\mathbf g}$, where
\begin{eqnarray}
   {\mathbf g}= \left[\begin{array}{ccc} {0}&{1}&{0}\\ {1}&{0}&{0}\\
	{0}&{0}&{1} \end{array}\right]\,.
\label{e4.7}
\end{eqnarray}
We note that there may exist other forms of ${\mathbf g}$  such that
$\mathbf G$ satisfies Eq.~(\ref{e4.6}),  but Eq.~(\ref{e4.7}) is the
simplest one that satisfies ${\mathbf G}^{2}\propto {\mathbf I}$ and
works for all $\gamma$.

For the purpose of future use, we construct matrices ${\mathbf T}_0$
and ${\mathbf H}_0(\Delta t)$ by setting $\gamma=0$ in
Eq.~(\ref{e4.1}) and Eq.~(\ref{e4.3}), without setting $\alpha_0=0$
(in reality, $\alpha_0=0$ when $\gamma=0$), i.e.,
\begin{eqnarray}
{\mathbf T}_0&\equiv&  \left[\begin{array}{cc} {\mathbf  0}&{{\mathbf
        I}}\\  {\mathbf 0} &{-\,\alpha_{0}{\mathbf I}}
        \end{array}\right]\quad\mbox{and}
\label{e4.8}\\
	{\mathbf H}_0(\Delta t) &\equiv& \exp\left[{\mathbf T}_0
\Delta t\right]
\label{e4.9}
\,.
\end{eqnarray}
More explicitly, the  $3\times3$ sub-blocks of ${\mathbf H}_0(\Delta
t)$ are given by  $({\mathbf H}_{0})^{[k]}_{lm}(\Delta t)={\mathbf
h}_{0}^{(k)}(\Delta t)\,\delta_{l,m}$, and ${\mathbf
h}_{0}^{(k)}(\Delta t)$ can be found by putting $\gamma=0$ in
Eq.~(\ref{e4.3}) without putting $\alpha_{0}=0$. The matrix ${\mathbf
H}_0(\Delta t)$ is now not only generalized  $\mu$-symplectic with
matrix $\mathbf G$, but also $\mu$-symplectic with $\mathbf J$, i.e.,
\begin{eqnarray}
    [{\mathbf{H}}_{0}(\Delta t)]^{\mbox{\scriptsize T}}\,{\mathbf
 J}\,{\mathbf{H}}_{0}(\Delta t)\, =\,e^{-\alpha_{0}\Delta t}{\mathbf
 J}\,.
\label{e4.10}
\end{eqnarray}
The relevance of this observation will become clear in Sec.~IV.C.

We also note that Eqs.~(\ref{e4.1}-\ref{e4.7}) hold for any constant
coefficient of friction (not necessarily $\alpha_{0}$), which implies,
using  Eq.~(\ref{e2.21}), that for a {\it collisionless\/} gas of
point particles obeying the SLLOD dynamics with a constant coefficient
of friction, the CPR is exact, as can be seen in simulations
\cite{SEI_chaos_98}.

\subsection{Generalized $\mu$-symplecticity Property of ${\mathbf
M}_{ij}$}

Unlike the  ${\mathbf H}$ matrices, the  ${\mathbf M}_{ij}$  matrices
corresponding to a binary collision between the $i$-th and the $j$-th
sphere do not follow from Eq.~(\ref{e2.9}) directly. This is due to
the fact that even though the sequence of binary collisions are the
same on the reference and the adjacent trajectories, the binary
collision between the $i$-th and the $j$-th sphere on these two
trajectories in the phase space are {\it not\/} simultaneous.   One
therefore needs the dynamics of the tangent vectors for the time
interval, $\delta\tau$, between the two collisions on the reference
and the adjacent trajectories involving the $i$-th and the $j$-th
sphere. To obtain  an expression of  ${\mathbf M}_{ij}$, we follow the
formalism developed in Refs.~\cite{Panja_preprint,vanZon_thesis},
which in turn, is based on the formalism presented in
Refs.~\cite{GaspardDorfman} and \cite{DPH_pre_96}.

The dynamics of tangent vectors at a collision is derived in the
Appendix.  The result is that
\begin{eqnarray}
	\vec{\delta\Gamma}_+ =  {\mathbf M}_{ij} \vec{\delta\Gamma}_-
\label{e4.11}
\end{eqnarray}
with
\begin{eqnarray} 
   {\mathbf M}_{ij} = ({\mathbf
I}-2\hat{N}_{ij}\hat{N}_{ij})\,\left[\begin{array}{cc} {{\mathbf
I}\,\,\,\,\,\,\,\,}&{\mathbf 0}\\ {\mathbf R\,\,\,\,\,\,\,}&{\mathbf
I}\!  \end{array}\right]\,,
\label{e4.12}
\end{eqnarray}
and ${\mathbf R}$ a symmetric matrix given by Eq.~(\ref{eaR}).  This
form of ${\mathbf M}_{ij}$  immediately implies  that ${\mathbf M}$ is
symplectic, but {\it not\/} generalized $\mu$-symplectic with matrix
${\mathbf G}$ for $\mu=1$, i.e.,
\begin{eqnarray}
   {\mathbf{M}}_{ij}^{\mbox{\scriptsize T}}{\mathbf
	J}{\mathbf{M}}_{ij}= {\mathbf J}\,
\label{e4.13}
\end{eqnarray}
but
\begin{eqnarray}
   {\mathbf{M}}_{ij}^{\mbox{\scriptsize T}}{\mathbf
G}{\mathbf{M}}_{ij} \,\neq\, {\mathbf G}\,.
\label{e4.14}
\end{eqnarray}

\subsection{Generalized $\mu$-symplecticity Property of ${\mathbf L}
(t-t_{0})$ and the Origin of an Approximate CPR}

{}From  Secs.~IV.A and IV.B above, we can finally see that for a
collection of hard spheres obeying the SLLOD equations of motion with
constant coefficient of friction $\alpha_{0}$
\begin{enumerate}
\item[(a)] the  ${\mathbf H}$ matrices are generalized
$\mu$-symplectic with matrix ${\mathbf G}$, but {\it not\/} with
matrix ${\mathbf J}$  (see Eq.~(\ref{e4.6}))
\item[(b)]  the ${\mathbf M}$ matrices are symplectic but {\it not\/}
generalized $\mu$-symplectic with matrix ${\mathbf G}$  (see
Eqs.~(\ref{e4.13}) and (\ref{e4.14})) .
\end{enumerate}
Once the ${\mathbf H}$ and the ${\mathbf M}$ matrices are combined
together, following Eq.~(\ref{e2.16}), the matrix
\begin{eqnarray}
{\mathbf L}(t-t_{0})\!={\mathbf H}(\Delta t_{s}){\mathbf M}_{i_sj_s}
	{\mathbf H}(\Delta t_{s-1})\ldots{\mathbf M}_{i_1j_1}{\mathbf
	H}(\Delta t_{0})
\label{e4.15}
\end{eqnarray}
is seen to be generalized $\mu$-symplectic {\it neither\,} with matrix
${\mathbf G}$ {\it nor\,} with matrix ${\mathbf J}$.

To study the degree of deviation from an exact CPR using the
properties of ${\mathbf L}(t-t_{0})$ in Eq.~(\ref{e4.15}) with
constant coefficient of friction $\alpha_{0}$, we can use either
${\mathbf K}={\mathbf G}$ or ${\mathbf K}={\mathbf J}$. While the
former choice implies that one has to try to estimate the deviation
from an exact CPR using the distribution of the unit vectors
$\hat{N}_{ij}$'s and the collision angles for different sets of binary
collisions in the expression of ${\mathbf M}$, the latter choice means
that one can make the estimate by using the  property of the ${\mathbf
H}$ matrices in Eq.~(\ref{e4.15}).  We choose the latter approach,
because not only is it much easier to calculate the typical magnitude
of a flight time of a sphere at low densities, but also, an estimate
of the deviation from the exact CPR can be made at small $\gamma$, as
a {\it power series expansion} in $\gamma$.  However, the smallness of
$\gamma$, which has a dimension of inverse time, has to be defined in
a proper manner. To do so, we notice that the density of the spheres
$n$ sets a time scale in the form of the mean flight time $\tau_{0}$
of an individual sphere, and in three dimensions $\tau_{0}\sim\tilde n
a_0/v_0$. Here,  $\tilde{n}=na_0^{3}$ is the dimensionless density
and, as before, $v_0$ and $a_0$ are typical velocity and radius of a
sphere.  Thus, the actual dimensionless small parameter corresponding
to the shear rate is $\tilde\gamma=\gamma\tau_{0}$.

A naive way to  estimate the deviation from an exact CPR using the
latter approach is to use the deviation of the ${\mathbf H}(\Delta t)$
matrices from an exact $\mu$-symplecticity (see
Eq.~(\ref{e4.15})). Such a deviation is characterized by the matrix
${\mathbf D(\Delta t)}=[{\mathbf{H}}(\Delta t)]^{\mbox{\scriptsize
T}}{\mathbf J}{\mathbf{H}}(\Delta t)-e^{-\alpha_{0}\Delta t}{\mathbf
J}$.  The matrix ${\mathbf D}(\Delta t)$  can easily be calculated
from Eq.~(\ref{e4.3}). However, to estimate the order of the matrix
elements of ${\mathbf D}(\Delta t)$,  an order of estimate of the
quantity  $\Delta t$ has to be obtained. To this end, we note that
while $\tau_{0}$ is the mean flight time for an {\it individual\/}
sphere,  $\Delta t$  in Eq.~(\ref{e4.15}) denotes the mean time for a
flight of $N$ spheres. This implies that $\Delta t\sim{\tau_{0}}/N$,
as on an average, there are $N/2$ different binary collisions over a
mean flight time $\tau_{0}$ of an individual sphere. Thus, one would
expect that in the thermodynamic limit,  ${\mathbf D}(\Delta
t)\rightarrow{\mathbf D}(0)={\mathbf 0}$  and one would be led to
conclude that  the ${\mathbf H}$ matrices in Eq.~(\ref{e4.15}) are all
symplectic. This in turn would imply, from Eq.~(\ref{e4.15}), that
${\mathbf L}(t-t_{0})$ would be  $\mu$-symplectic and therefore a gas
of hard spheres, obeying the SLLOD dynamics with a constant
coefficient of friction $\alpha_{0}$ would satisfy an exact CPR in the
thermodynamic limit. We demonstrate below that this simplification is
{\it not\/} correct.

The proper estimate of the deviation from an exact CPR has to be made
by considering ${\mathbf H}(\tau_{0})$. To see why this is so, we
rewrite the matrix ${\mathbf H}(\Delta t)$ as
\begin{eqnarray}
   {\mathbf H}(\Delta t)= e^{\frac{N-1}{2}\,\alpha_{0}\Delta
	t}\,\prod_{i=1}^N{\mathbf H}^{i}(\Delta t)
\label{e4.16}
\end{eqnarray}
with  ${\mathbf H}^{i}(\Delta t)$ defined by
\begin{eqnarray}
   ({\mathbf H}^i)^{[k]}_{lm}(\Delta t) &=&
	\delta_{l,m}\big[\,\delta_{l,i}{\mathbf h}^{(k)}(\Delta t)
	\nonumber \\&& +(1-\delta_{l,i})(\delta_{k,1}
	+\delta_{k,4})e^{-\alpha_0\Delta t/2}\,{\mathbf I}\,\big]\,.
\label{e4.17}
\end{eqnarray}
In effect, ${\mathbf H}^{i}(\Delta t)$ describes the evolution of the
infinitesimal deviation of the trajectory of the $i$-th  sphere, while
it has an almost trivial action on the infinitesimal deviation of the
trajectory of the  $j$-th sphere, $j\neq{i}$. It is easy to see that
${\mathbf H}^{i}(\Delta t)$ has some useful properties
\begin{eqnarray}
   {\mathbf H}^{i}(\Delta t_{1}) \,{\mathbf H}^{i}(\Delta t_{2})&=&
	{\mathbf H}^{i}(\Delta t_1+\Delta t_2)\,, \nonumber \\
	\,[\,{\mathbf H}^{i}(\Delta t_{i}),\,{\mathbf H}^{j}(\Delta
	t_j)\,] &=& 0 \quad\mbox{and} \nonumber \\ \,[\,{\mathbf
	H}^{i}(\Delta t),\,{\mathbf M}_{i_{c}\,j_{c}}\,] &=&
	0\quad\mbox{if $i_{c}\neq i$ and $j_{c}\neq i$}\,.
\label{e4.18}
\end{eqnarray}

The properties of ${\mathbf H}^{i}(\Delta t)$  in Eq.~(\ref{e4.18})
allow us to shuffle the terms in Eq.~(\ref{e4.15}) such as to collect
together as many  ${\mathbf H}^{i}(\Delta t)$s with the same $i$ as
possible. The result is that to the right of any ${\mathbf
M}_{i_cj_c}$ figure an  ${\mathbf H}^{i_c}(\tau_{i_c}^c)$ and an
${\mathbf H}^{j_c}(\tau_{j_c}^c)$, where $\tau_{i_c}^c$ and
$\tau_{j_c}^c$ are the time of flights for the $i_{c}$-th and the
$j_{c}$-th  spheres before their mutual collision $c$. Consequently,
\begin{eqnarray}
   {\mathbf L}(t-t_{0})&=& e^{\frac{N-1}{2}\,\alpha_{0}(t-t_0)}
	\left\{\prod_{i=1}^{N} {\mathbf H}^{i}(t-\tilde t_i)\right\}
	\nonumber \\ && \times\prod_{c=1}^{s}{\mathbf M}_{i_c\,j_c}\,
	{\mathbf H}^{i_c}(\tau^c_{i_c})\,{\mathbf
	H}^{j_c}(\tau^c_{j_c}).
\label{e4.19}
\end{eqnarray}
The product sign in Eq.~(\ref{e4.19}) is to be expanded towards the
left, i.e.  $\prod_{c=1}^s A_c=A_s\cdots A_1$. Here, $\tilde t_i$ is
the last time that  $i$-th particle collided (or $t_0$ if it didn't
collide). From Eq.~(\ref{e4.19}), it is now clear that the  proper
estimate for the deviation from an exact CPR has to be made by
considering the properties of  ${\mathbf H}^{i}(\tau_i)$, with
$\tau_i=O(\tau_0)$, and {\it not\/} from the properties of ${\mathbf
H}^{i}(\tau_0/N)$.

We also notice that  if one uses the corresponding ${\mathbf H}_{0}$
matrices  instead of the ${\mathbf H}$ matrices in Eq.~(\ref{e4.15})
to construct an analogous matrix ${\mathbf L}_{0}(t-t_{0})$, defined by
\begin{eqnarray}
{\mathbf L}_{0}(t-t_{0}) &=& {\mathbf H}_{0}(\Delta t_{s})\,{\mathbf
	M}_{i_sj_s}\, {\mathbf H}_{0}(\Delta t_{s-1}) \ldots\nonumber
	\\&&  \ldots {\mathbf M}_{i_1j_1}\,{\mathbf H}_{0}(\Delta
	t_{0})\,,
\label{e4.20}
\end{eqnarray}
then the matrix ${\mathbf L}_{0}(t-t_{0})$ is $\mu$-symplectic,
because of  Eqs.~(\ref{e4.10}) and (\ref{e4.13}). As a consequence,
the logarithms of the eigenvalues of ${\mathbf\tilde{L}}_0(t-t_{0})$,
defined by
\begin{eqnarray}
{\mathbf\tilde{L}}_{0}(t-t_{0})\,=\,[{\mathbf
L}_{0}(t-t_{0})]^{\mbox{\scriptsize T}}{\mathbf L}_{0}(t-t_{0})\,,
\label{e4.21}
\end{eqnarray}
pair exactly. If we arrange the corresponding Lyapunov spectrum
\begin{eqnarray}
 {\mathbf \Lambda}_0 &=& \lim_{t\rightarrow\infty}\left[
        {\mathbf\tilde{L}}_0(t-t_{0})\right]^{1/2(t-t_{0})}\,,
\label{e4.22}
\end{eqnarray}
in the decreasing order of magnitude as $\lambda^{(0)}_{1}$,
$\lambda^{(0)}_{2},$ $\ldots,\lambda^{(0)}_{6N}$, then
$\lambda^{(0)}_{i}+\lambda^{(0)}_{6N-i+1}=-\alpha_{0}$.
  
Motivated by  this, our approach to study the deviation from an exact
CPR for the matrix ${\mathbf L}(t-t_{0})$ will be to take ${\mathbf
L}_{0}(t-t_{0})$ as the reference matrix. This  is reasonable because,
as we will show, ${\mathbf L}(t-t_{0})$ and ${\mathbf L}_{0}(t-t_{0})$
are very close for small $\tilde\gamma$, if
$t-t_{0}=\tau=O(\tau_{0})$. To show this, we first write the matrix
${\mathbf L}_{0}(t-t_{0})$ in the same form as Eq.~(\ref{e4.19}), with
${\mathbf H}^{i}(\tau_i)$s replaced by ${\mathbf
H}^{i}_{0}(\tau_i)$s. We then relate, for $t-t_{0}=\tau=O(\tau_{0})$,
the difference matrix ${\mathbf\Delta{L}}(t-t_{0})={\mathbf
L}(t-t_{0})-{\mathbf L}_{0}(t-t_{0})$ to the difference between
${\mathbf H}^i(\tau_i)$ and ${\mathbf H}_0^i(\tau_i)$.
\label{anticipate}

We now define new matrices $\Delta {\mathbf H}^i$ and ${\mathbf E}^i$,
such that
\begin{eqnarray}
\Delta {\mathbf H}^i(\tau_i) &\equiv& {\mathbf H}^i(\tau_i)-{\mathbf
H}_0^i(\tau_i) \nonumber\\ {\mathbf E}^i(\tau_i) &\equiv & \left[
{\mathbf H}_0^i(\tau_i) \right]^{-1} \Delta{\mathbf H}^i(\tau_i)
\label{e4.23}
\end{eqnarray}
with
\begin{eqnarray}
   ({\mathbf H}^i_0)^{[k]}_{lm}(\tau_i) &=&
	\delta_{l,m}\big[\,\delta_{l,i}{\mathbf h}_0^{(k)}(\tau_i)
	\nonumber \\&& +(1-\delta_{l,i})(\delta_{k,1}
	+\delta_{k,4})e^{-\alpha_0\tau_i/2}\,{\mathbf I}\,\big]\,
\label{e4.23a}
\end{eqnarray}
[for the definition of ${\mathbf h}_0^{(k)}(\tau_i)$, see the
preceding paragraph of Eq.~(\ref{e4.10})]. It is easily seen that
${\mathbf E}^i(\tau_i)$ has non-zero elements  only for entries
involving  the $i$-th sphere. This, together with relations
(\ref{e4.18}), implies that
\begin{eqnarray}
\,[\,{\mathbf E}^{i}(\tau_{i}),\,{\mathbf H}^{j}_0(\tau_j)\,] &=& 0
\quad\mbox{for $i\neq j$ and} \nonumber\\ \,[\,{\mathbf
E}^{i}(\tau_i),\,{\mathbf M}_{i_{c}\,j_{c}}\,] &=& 0\quad\mbox{if
$i_{c}\neq i$ and $j_{c}\neq i$}\,.
\label{e4.24}
\end{eqnarray}

The matrix ${\mathbf E}^{i}(\tau_i)$ can be easily calculated from
Eqs.~(\ref{e4.23}). With the aid of Eqs.~(\ref{e4.3}) and
(\ref{e4.9}), we now express the matrix $\Delta {\mathbf H}^i(\tau_i)$
defined in Eq.~(\ref{e4.23}) as $\Delta {\mathbf
H}^{i(k)}_{lm}(\tau_i)=\Delta{\mathbf
h}^{i[k]}(\tau_i)\,\delta_{l,m}$, where
\begin{eqnarray}
   \Delta{\mathbf h}^{i(1)}(\tau_i)&=&\left[\,\gamma\tau_i -{
\frac{\gamma\,[1-\exp(-\,\alpha_{0}\tau_i)]}
{\alpha_{0}}}\right]{\mathbf c}\,,  \nonumber \\  \Delta{\mathbf
h}^{i(2)}(\tau_i)&=&  \frac{\gamma}{\alpha_0^2}\left[ \alpha_0\tau_i
(1+e^{-\alpha_0\tau_i})-2+2e^{-\alpha_0\tau_i} \right] {\mathbf c} \,,
\nonumber \\  \Delta{\mathbf h}^{i(3)}(\tau_i)&=&
\gamma\,[1-e^{-\alpha_{0}\tau_i}]\,{\mathbf c}\,\quad\mbox{and}
\nonumber \\ \Delta{\mathbf h}^{i(4)}(\tau_i )&=& -\gamma
e^{-\alpha_0\tau_i} \left[\tau_i +{\frac{1}
{\alpha_{0}}[1-e^{\alpha_{0}\tau_i}]}\right] {\mathbf c}\,.
\label{add1}
\end{eqnarray} 
The crucial point is that since $\alpha_0\propto\gamma\tilde\gamma$,
i.e., the dissipation is quadratic in the shear field,  $\Delta
{\mathbf H}^i(\tau_i)$ in Eq. (\ref{add1}) can be expanded in powers
of $\gamma$ to obtain
\begin{eqnarray}
    \Delta{\mathbf
	h}^{i(1)}(\tau_i)&=&\frac{\gamma\alpha_0\tau^2_i}{2!}
	\,{\mathbf c}\,, \nonumber \\ \Delta{\mathbf
	h}^{i(2)}(\tau_i)&=&  \frac{\gamma\alpha_0\tau^3_i}
	{3!}\,{\mathbf c}\,, \nonumber \\  \Delta{\mathbf
	h}^{i(3)}(\tau_i)&=& \gamma\alpha_{0}\tau_i\,{\mathbf
	c}\,\quad\mbox{and} \nonumber \\ \Delta{\mathbf
	h}^{i(4)}(\tau_i )&=&\frac{\gamma\alpha_0\tau^2_i}{2!}
	\,{\mathbf c}
\label{add2}
\end{eqnarray} 
at the {\it most leading order.} Thereafter, having combined
Eqs.~(\ref{e4.23}) and (\ref{add1}-\ref{add2}), {\it all\/} elements
of ${\mathbf  E}^i(\tau_{i})$ are seen to have a prefactor of order
$\tilde\gamma^3$.  This makes ${\mathbf E}^i(\tau_{i})$ small compared
to $1$. Hence, to first order, ${\mathbf \Delta L}(\tau)={\mathbf
L}(\tau)-{\mathbf L}_0(\tau)$ can be found from Eq.~(\ref{e4.19}) by
keeping only  the terms  linear in  ${\mathbf E}^i(\tau_i)$: \bleq
\begin{eqnarray}
     \Delta {\mathbf L}(\tau) &=&  e^{\frac{N-1}{2}\alpha_0\tau}
	\left\{\prod_{i=1}^{N} {\mathbf H}^{i}_{0}(\tau+t_0- \tilde
	t_i)\right\} \Bigg( \sum_{j=1}^N  {\mathbf
	E}^j(\tau+t_0-\tilde t_j) \prod_{c=1}^{s}{\mathbf M}_{c}\,
	{\mathbf H}^{i_c}_{0}(\tau^c_{i_c})\,{\mathbf
	H}^{j_c}_{0}(\tau^c_{j_c})  \nonumber\\&& + \sum_{c=1}^{s}
	\prod_{a=1}^{c}  {\mathbf M}_{a} {\mathbf
	H}_0^{i_a}(\tau_{i_a}^a) {\mathbf H}_0^{j_a}(\tau_{j_a}^a)
	\times \left[ {\mathbf E}^{j_c}(\tau_{j_c}^c) +{\mathbf
	E}^{i_c}(\tau_{i_c}^c)  \right] \prod_{b=c+1}^{s}  {\mathbf
	M}_{b}{\mathbf H}_0^{i_b}(\tau_{i_b}^b) {\mathbf
	H}_0^{j_b}(\tau_{j_b}^b) \Bigg)\,.
\label{e4.25}
\end{eqnarray}
We now shuffle all the  ${\mathbf E}^{i_c}$ matrices to the right, by
successively exchanging the order of ${\mathbf E}^{i_c}$  and the next
${\mathbf M}_{b}{\mathbf H}_0^{i_b}(\tau_{i_b}^b){\mathbf
H}_0^{j_b}(\tau_{j_b}^b)$ to its right.  These will commute very
often, because most collisions in the sequence will involve other
spheres than this $i_c$-th one.  If they do not commute, we write
\begin{eqnarray}
	{\mathbf E}^{i_c} {\mathbf M}_{b} {\mathbf
	H}_0^{i_b}(\tau_{i_b}^b)  {\mathbf H}_0^{j_b}(\tau_{j_b}^b) =
	{\mathbf M}_{b}{\mathbf H}_0^{i_b}(\tau_{i_b}^b) {\mathbf
	H}_0^{j_b} (\tau_{j_b}^b) \left[{\mathbf M}_{b}{\mathbf
	H}_0^{i_b}(\tau_{i_b}^b) {\mathbf H}_0^{j_b}
	(\tau_{j_b}^b)\right]^{-1}\!\!   {\mathbf E}^i {\mathbf
	M}_{b}{\mathbf H}_0^{i_b}(\tau_{i_b}^b) {\mathbf
	H}_0^{j_b}(\tau_{j_b}^b)\,.
\label{e4.26}
\end{eqnarray}
From that point on, we work with  ``$[{\mathbf M}_{b}{\mathbf
H}_0^{i_b}{\mathbf H}_0^{j_b}]^{-1}$  ${\mathbf E}^i {\mathbf
M}_{b}{\mathbf H}_0^{i_b} {\mathbf H}_0^{j_b}$'', and shuffle that to
the right.  Repeating this, we end up with ${\mathbf L}_0(\tau)$ on
the left side again, and find, symbolically,
\begin{eqnarray}
\left[{\mathbf L}_0(\tau)\right]^{-1} \Delta {\mathbf L}(\tau)&=&
	\sum_{c=1}^s   \left\{	\prod({\mathbf MH_0H_0})^{-1} \right\}
	\left[ {\mathbf E}^{j_c}(\tau_{j_c}^c) +{\mathbf
	E}^{i_c}(\tau_{i_c}^c) \right]  \left\{\prod{\mathbf MH_0H_0}
	\right\} \nonumber\\ && +\sum_{j=1}^N \left\{\prod({\mathbf
	MH_0H_0})^{-1}  \right\} {\mathbf E}^j(\tau+t_0-\tilde t_j)
	\left\{\prod{\mathbf MH_0H_0}  \right\} \,.
\label{e4.27}
\end{eqnarray}
\eleq
\noindent
Since $\tau$ is of the order of a  flight time of a sphere, $s$ is of
the order of $N$, but a typical sphere will have suffered only a few
[a few here is roughly $O(1)$] collisions. Consider a typical term in
the sum over $c$. The ${\mathbf MH_0H_0}$ terms in the products are
from collisions that involve  the $i_c$-th or the $j_c$-th sphere, or
spheres that were involved in collisions with the $i_c$-th or $j_c$-th
sphere.  The number of ${\mathbf MH_0H_0}$ will thus also be of $O(1)$.

The matrix ${\mathbf E}^{i_c}$ has non-zero elements only for the
entries associated with the $i_c$-th  sphere. The multiplication by
${\mathbf MH_0H_0}$ yields more non-zero entries associated with other
spheres, whose number however is of $O(1)$. Therefore, in the whole
sum of $s=O(N)$ terms in Eq.~(\ref{e4.27}) will yield for typical
entries a result of order one (and {\it not\,} of order $N$ as the sum
over $s\propto N$ suggests).  Since {\it all\/} elements of the
${\mathbf E}$ matrices are proportional to $\gamma\tilde\gamma^2$ as
can be seen from Eqs.~(\ref{e4.23}), (\ref{add1}) and (\ref{add2}), we
conclude for $t-t_0=\tau=O(\tau_0)$ that
\begin{eqnarray}
	{\mathbf L}(\tau) = {\mathbf L}_0(\tau)\,[{\mathbf
	I}\,+\,\tilde\gamma^3 {\mathbf B}\,]\,,
\label{e4.28}
\end{eqnarray}
where the matrix ${\mathbf B}$ is of order 1 in $\tilde\gamma$  and
order 1 in $N$.  Note that $\mathbf B$ contains higher powers of
$\tilde\gamma$ as well.  Because it involves $\mathbf c$ and
contributions from collisions between spheres,  ${\mathbf B}$ is not
proportional to the identity $\mathbf I$, so we cannot regard it as a
simple scalar factor (in which case  the exact conjugate pairing would
be easy to obtain again). Equation (\ref{e4.28}) implies
\begin{eqnarray}
   {\mathbf\Delta\tilde L}(\tau)&=&{\mathbf \tilde
L}(\tau)\,-\,{\mathbf \tilde L}_0(\tau) \nonumber \\&=&
\tilde\gamma^3\,[\,{\mathbf B}^{\mbox{\scriptsize T}}{\mathbf \tilde
L}_0(\tau)\,+\,{\mathbf \tilde L}_0(\tau){\mathbf B}\,]
\,+\,\tilde\gamma^6{\mathbf B}^{\mbox{\scriptsize T}}{\mathbf \tilde
L}_0(\tau){\mathbf B}.
\label{e4.29}
\end{eqnarray}
We can now see, from Eqs.~(\ref{e4.28}) and (\ref{e4.29}) that the
differences between ${\mathbf L}(\tau)$ and ${\mathbf L}_0(\tau)$, and
between ${\mathbf \tilde L}(\tau)$ and ${\mathbf \tilde L}_0(\tau)$
are small, as anticipated on page \pageref{anticipate},  by a relative
order $\tilde\gamma^3$.  Therefore the logarithm of the eigenvalues of
${\mathbf L}(\tau)$ and ${\mathbf L}_0(\tau)$ also differ by a term of
order $\tilde\gamma^3$ in an absolute sense.  If we now divide the
logarithms of these eigenvalues by the time $\tau$,  we see that the
finite time  (for time $\tau$) Lyapunov exponents, calculated from
${\mathbf\tilde L}_0$ and from ${\mathbf \tilde L}$  (which we denote
as $\lambda^{(0)}_{i}(\tau)$ and $\lambda_{i}(\tau)$ respectively, for
$i=1,2\ldots6N$) differ by a term
$O(\tilde\gamma^3/\tau)=O(\gamma\tilde\gamma^2)$.  Using the fact that
that  $\lambda^{(0)}_{i}(\tau)$s satisfy the conjugate pairing rule,
$\lambda_{i}(\tau)$s too, will satisfy it up to
$O(\gamma\tilde\gamma^2)$. From  the condition that
\begin{eqnarray}
{\sum_{i=1}^{6N}}\lambda^{(0)}_{i}(\tau)={\sum_{i=1}^{6N}}\lambda_{i}(\tau)
=-3N\alpha_{0}\,,
\label{e4.29a}
\end{eqnarray}
[which can be easily verified from Eqs.~(\ref{e4.3}) and
(\ref{e4.12})],  we see that on average over the pairs, the deviation
from CPR is zero.

We make one further observation at this stage.  The Lyapunov exponents
(even the finite time once) are invariant under
$\gamma\rightarrow-\gamma$, so in a power series expansion of the
individual Lyapunov exponents in $\tilde\gamma$ \cite{analytic},  the
odd powers vanish. We therefore conclude that the logarithm of the
eigenvalues of ${\mathbf L}(\tau)$ and ${\mathbf L}_0(\tau)$ must
differ by a term of order $\tilde\gamma^4$, and hence conjugate
pairing of $\lambda_{i}(\tau)$s  will be valid up to the correction of
the form $\gamma\tilde\gamma^3$.

To extend these ideas for large $(t-t_{0})$ [and finally for
$t\rightarrow\infty)$], we proceed in the following way. Notice that
the matrices ${\mathbf\tilde L}(t-t_0)$ and ${\mathbf\tilde
L}_{0}(t-t_0)$ are positive definite and symmetric. This allows us to
express the two matrices in the form of ${\mathbf\tilde
L}_{0}(t-t_0)=\exp({\mathbf A}_{0})$ and ${\mathbf\tilde
L}(t-t_0)=\exp({\mathbf A})$, where for large $(t-t_{0})$, both the
eigenvalues of ${\mathbf A}_{0}$ and ${\mathbf A}$ must behave
$\sim(t-t_{0})$. In these terms, the difference between the Lyapunov
exponents for ${\mathbf\tilde L}(t-t_0)$ and ${\mathbf\tilde
L}_{0}(t-t_0)$ is related to ${\mathbf A}-{\mathbf A}_{0}$. Since the
difference between ${\mathbf\tilde L}(t-t_0)$ and ${\mathbf\tilde
L}_{0}(t-t_0)$ has an explicit prefactor of $\tilde\gamma^{3}$, so
will ${\mathbf A}-{\mathbf A}_{0}$. Using the symmetry argument that
the Lyapunov exponents have to be even functions of $\gamma$, we obtain
\begin{eqnarray}
\lambda_i +  \lambda_{6N-i+1}\,=\, -\alpha_0 +
O(\gamma\tilde\gamma^3)\,.\quad i=1,\ldots6N
\label{e4.30}
\end{eqnarray}
To explicitly extend the  formalism developed in
Eqs. (\ref{e4.19}-\ref{e4.29}) to  large $t-t_{0}$ and  thereby obtain
a relation between $\lambda_{i}$s and $\lambda^{(0)}_{i}$s,  we need
to concatenate a lot of ${\mathbf L}(\tau)$s. These do not commute
with each other, nor do they commute with ${\mathbf B}$s in
general. This prevents  an explicit demonstration of how the deviation
is built up.  For the largest and the most negative Lyapunov exponent,
it has been possible to show that they pair to $-\alpha_0$ plus
corrections of $O(\gamma\tilde\gamma^3)$  by means of a kinetic theory
approach \cite{vanZon_thesis,vZvB_unpublished}, based on the
independence of subsequent collisions of a  sphere. One expects that
in subsequent time-intervals of $O(\tau_0)$, the ${\mathbf L}(\tau)$
matrices are not qualitatively much different from each other.
Therefore, we expect  that the coefficient of the
$O(\gamma\tilde\gamma^3)$  term in Eq.~(\ref{e4.30}), to be of the
same order as that for a flight  time $\tau=O(\tau_{0})$, i.e. of the
order of ${\mathbf B}$, which is $O(1)$.

We now return to the discussion of an approximate CPR for the SLLOD
equations of motion with an iso-kinetic Gaussian thermostat in between
collisions.  As discussed in Sec.~III, in the non-equilibrium steady
state and in the thermodynamic limit, the coefficient of dynamical
friction representing the iso-kinetic Gaussian thermostat for the
SLLOD dynamics fluctuates with $1/\sqrt{N}$ fluctuations around the
fixed value $\alpha_{0}$. We would therefore expect that the Lyapunov
spectrum for the SLLOD dynamics with an iso-kinetic Gaussian
thermostat are given by that of ${\mathbf L}(t-t_0)$ {\it plus\,}
terms of $O(1/\sqrt{N})$. Consequently, the approximate conjugate
pairing of Eq.~(\ref{e4.30}) can  be extended to this thermostat once
we neglect the $O(1/\sqrt{N})$ terms in the sum of the corresponding
$\lambda_i$ and $\lambda_{6N-i+1}$.

\section{Discussion}

In this paper, we started with a collection of $N$ hard spheres, each
with unit mass and arbitrary radius. Next, we argued how the
coefficient of friction representing the iso-kinetic Gaussian
thermostat in between collisions, fluctuates around a fixed value
$\alpha_{0}$ with $1/\sqrt{N}$ fluctuations in the NESS. Using the
properties of the transformation matrices for the infinitesimal phase
space element $\vec{\delta\Gamma}$, we then showed that the CPR is
expected to be violated, at the most, at $O(\gamma^{4})$ for constant
coefficient of friction $\alpha_0$.  The source of the CPR violation
is basically the $\alpha_0\gamma{\mathbf C}\vec{R}$ term in
Eq. (\ref{e2.7}), as it is clearly seen that CPR would have been exact
in the absence of this term.  Finally, we extended that result to the
case when the coefficient of friction represents an iso-kinetic
Gaussian thermostat. {\it Note that to obtain the deviations from  CPR
for the constant $\alpha_0$ thermostat, the number of particles  did
not matter, so for that case the result holds for any number of
particles, but with an iso-kinetic Gaussian thermostat between
collisions, we need to take $N$ to infinity to make the connection  to
the constant $\alpha_0$ thermostat\/}.

In addition, as mentioned in the introduction of this paper, the
condition that the mass of each sphere be unity, is not necessary. If
we assume that the mass of the $i$-th sphere is $m_i$, then one can
obtain the same equations of motion, Eq.~(\ref{e2.1}-\ref{e2.2}) and
(\ref{e2.7}-\ref{e2.9}) in terms of the primed quantities defined by
$\vec{r}'_i=\sqrt{m_i}\vec{r}$, $\vec{v}'_i=\sqrt{m_i}\vec{v}$,
$\vec{p}'_i=\vec{p}_i/\sqrt{m_i}$,
$\vec{R}'=(\vec{r}'_{1},\vec{r}'_{2},...,\vec{r}'_{N})$,
$\vec{V}'=(\vec{v}'_{1},\vec{v}'_{2},...,\vec{v}'_{N})$ and
$\hat{N}'_{ij}$, defined as
\[
\hat{N}^{'\,l}_{ij}=
\left(\delta_{l,i}\sqrt{\frac{m_j}{m_i+m_j}}-\delta_{l,j}
\sqrt{\frac{m_i}{m_i+m_j}}\,\right)\,\hat{n}_{ij}
\]
$(l=1,2,..,N)$.  It is then straightforward, albeit lengthy and
laborious, to see that our entire analysis goes through in terms of
these primed variables, once we use ${\mathbf U}'$ in Eq.~(\ref{ea5})
with
\begin{eqnarray}
 {\mathbf U}'_{ij}&=&{\mathbf U}'_{ji} = [m_i+m_j]^{-1/2} \,{\mathbf
 I},\\ {\mathbf U}'_{ii} &=&  \frac{m_{j}}{m_{i}}{\mathbf U}'_{jj} =
 -\sqrt{\frac{m_{j}}{m_{i}}} {\mathbf U}'_{ij} \,.
\end{eqnarray}
However, we must note that even though the analysis presented in
Secs.~IV is not affected when the masses and the radii of the spheres
are not necessarily the same, one should not allow extreme variations
in masses and the radii of the spheres. For large variation of the
masses, the system may phase separate into phases in which the mean
flight times are different, which will invalidate the use of the
Boltzmann equation in Ref. \cite{Zon_pre_99} (the Boltzmann equation
has been used to show the approach of $\alpha$ towards the constant
value $\alpha_{0}$ in Ref. \cite{Zon_pre_99}). Furthermore, the use of
a typical mean free time in Sec. IV.B will not be possible for large
variation in the radii of the spheres. Our analysis in this paper
therefore, holds for somewhat limited variation of the masses and the
radii of the spheres.

Another possible generalization of our analysis can be carried out for
the case when the gas particles interact with each other by means of a
short-ranged, repulsive, inter-particle potential (an attractive
potential may cause bound states) for a constant multiplier
thermostat.  The dynamics of the particles can be again decomposed
into flights and ``collisions'' at low density. While the
transformation of the infinitesimal phase space volume element over
flights will again be determined by the ${\mathbf H}$ matrices as in
Eqs.~(\ref{e4.2}-\ref{e4.3}) (and thereby have the same properties as
above), the corresponding  ${\mathbf M}_{ij}$ matrices will not
necessarily  have a similar form and properties as presented in
Eqs.~(\ref{e4.12}-\ref{e4.13}). For a given average value of the
peculiar kinetic energy, the constant multiplier thermostat has  to be
chosen carefully (see the last paragraph of Sec. III).  Having denoted
the constant coefficient of friction again by  $\alpha_0$, where
$\alpha_0\propto\gamma^2$, is easy to see that the resulting matrix
${\mathbf M}_{ij}$  can be decomposed into a sum of two matrices,
${\mathbf M}^{(0)}_{ij}$ and ${\mathbf\Delta M}_{ij}$, where ${\mathbf
M}^{(0)}_{ij}$  is exactly $\mu$-symplectic but ${\mathbf\Delta
M}_{ij}$ is not.  If the term $\alpha_0\gamma{\mathbf C}\vec{R}$ in
Eq.~(\ref{e2.7}) (which is the culprit for the violations of CPR in
the first place) had been absent, then ${\mathbf{M}}_{ij}$ would have
been $\mu$-symplectic \cite{Dressler} and ${\mathbf \Delta M}_{ij}$
would have been zero, but with Eq.~(\ref{e2.7}) in its present form,
it is easily seen that ${\mathbf \Delta M}_{ij}$ is
$\propto\alpha\gamma\sim\tilde\gamma^{3}$ and the proportionality
constant depends on the ratio of the time the two colliding particles
spend in contact with each other to the time of a flight.  This ratio
is very small at low density of the gas and therefore the elements of
${\mathbf \Delta M}_{ij}$ are small in comparison with ${\mathbf
M}^{(0)}_{ij}$.  Since the ${\mathbf H}$ matrices have a similar
property, one can repeat the analysis of this paper for such
inter-particle short-range repulsive potentials using ${\mathbf H}_0$
and ${\mathbf M}^{(0)}_{ij}$ as the reference matrices and arrive at
the same conclusion as Eq.~(\ref{e4.30}).

Moreover, on the basis of the fact that at low densities and
short-ranged repulsive interparticle potentials (that do not give rise
to bound states), the constant multiplier thermostat and an
iso-kinetic Gaussian thermostat are equivalent in the thermodynamic
limit \cite{Zon_pre_99}, one would expect that under such conditions,
the Lyapunov exponent spectrum for the Gaussian thermostat would be
practically the same as that of the corresponding constant multiplier
thermostat, where the constant multiplier is once again proportional
to $\gamma^2$.  Therefore, in view of the discussion in the previous
paragraph,  we can also expect $O(\tilde\gamma^4)$ deviations from the
CPR for a Gaussian thermostatted SLLOD equations of motion in the
thermodynamic limit, where the particles interact with each other by
means of a short-ranged repulsive potential.

Finally, we note here that verifications of our theory by means of
computer simulation remains a challenge. The work for simulations  is
in progress at present, although our preliminary experience suggests
that to retrieve the  $\gamma\tilde\gamma^3$ scaling of the deviation
of the sum of the pairs of Lyapunov exponents from $-\alpha_0$ is not
an easy task.

\section*{Acknowledgment}

The authors wish to thank Prof. J. R. Dorfman and Prof. H. van
Beijeren for many useful and motivating discussions regarding this
subject. D.\ P. was supported  by the research grants from
``Fundamenteel Onderzoek der Materie (FOM)'' and that of Prof. J.\ R.\
Dorfman, NSF-PHY-9600428, and wishes to acknowledge the kind
hospitality of Prof.\ H.\ van Beijeren at University of Utrecht during
this and related work. R. v.\ Z. was supported by the research grant
of Prof. H. van Beijeren, by FOM, SMC and NWO Priority Program
Non-Linear Systems, which are financially supported by the
''Nederlandse Organisatie voor Wetenschappelijk Onderzoek (NWO)'', and
by a grant from the Natural Sciences and Engineering Research Council
of Canada.

\appendix

\section*{}

\setcounter{equation}{0}

\begin{figure}[h]
\begin{center}
\includegraphics[width=0.32\textwidth]{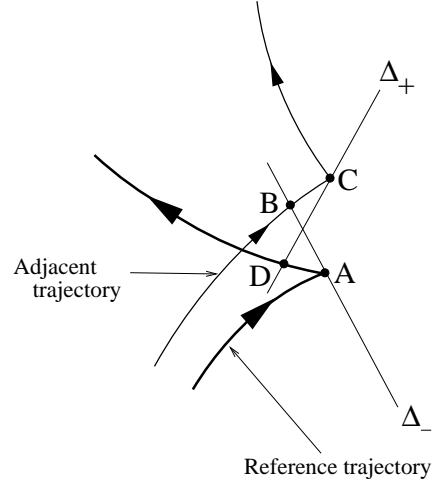}
\caption{A schematic diagram of the collision dynamics on the
reference and the adjacent trajectories.\label{fig3}}
\end{center}
\end{figure}

In this appendix, we will derive the transformation ${\mathbf M}_{ij}$
of tangent vectors at a collision. The essential element is that the
collision does not happen at the same time on reference and on the
adjacent trajectory.  To understand the origin of this time lag
between the binary collisions between the $i$-th and the $j$-th sphere
on the reference and the adjacent trajectories,  in Fig.~\ref{fig3},
we have depicted  an exaggerated schematic picture of the collisions
taking place in the $6N$-dimensional phase space on the reference and
the adjacent trajectories at points A and C respectively. The points B
and D show the corresponding positions of the adjacent and the
reference points respectively when the binary collisions at A and C
are taking place. Thus, the pre-collisional separation between the
reference and the adjacent points is
$\overrightarrow{\mbox{AB}}=\vec{\delta\Gamma}_{-}$, while the
post-collisional separation is
$\overrightarrow{\mbox{DC}}=\vec{\delta\Gamma}_{+}$. Using that $|\vec
r_j-\vec r_i|$ and  $|\vec r_j+\vec{\delta r}_j+\vec v_j\delta\tau
-\vec r_i-\vec{\delta r}_i-\vec v_i\delta\tau|$  both have to equal
$a_i+a_j$, the time lag  $\delta\tau$ between the two collisions at A
and C, therefore, can be easily expressed as
\begin{eqnarray}
   \delta\tau= -\frac{(\vec{\delta r}_{j-}-\vec{\delta
	r}_{i-})\cdot\hat n_{ij}}
	{(\vec{v}_{j-}-\vec{v}_{i-})\cdot\hat{n}_{ij}} =
	-\,\frac{\vec{\delta R}_{-}\cdot\hat{N}_{ij}}
	{\vec{V}_{-}\cdot\hat{N}_{ij}}\,.
\label{ea4.12}
\end{eqnarray}

To obtain the expression of $\vec{\delta\Gamma}_{+}$, we first express
the transformation of $(\vec{R},\vec{V})$ in Eq.~(\ref{e2.9}) over a
binary collision between the $i$-th and the $j$-th sphere in a matrix
form
\begin{eqnarray}
   \vec{\Gamma}_{+}= {\mathbf Q}\,(\vec{\Gamma}_{-}) =
	\left[\begin{array}{c}\vec{R}_{-}\\
	\vec{V}_{-}-2\,(\vec{V}_{-}\cdot\hat{N}_{ij})\,\hat{N}_{ij}
	\end{array} \right]\,,
\label{ea4.13}
\end{eqnarray}
Next we notice that
\begin{eqnarray}
   \vec{\delta\Gamma}_{+}= \frac{\partial{\mathbf Q}}
	{\partial\vec{\Gamma}_{-}}\cdot\vec{\delta\Gamma}^{*}_{-}
	-\dot{\vec{\Gamma}}_{+}\,\delta\tau\,,
\label{ea4.14}
\end{eqnarray}
where, $\vec{\delta\Gamma}^{*}_{-}$ is the infinitesimal phase space
separation between the two trajectories at A and C is
\begin{eqnarray}
   \vec{\delta\Gamma}^{*}_{-}=
	\vec{\delta\Gamma}_{-}\,+\,\dot{\vec{\Gamma}}_{-}\,\delta\tau
\label{ea4.15}
\end{eqnarray}

\begin{figure}[h]
\begin{center}
\includegraphics[width=0.4\textwidth]{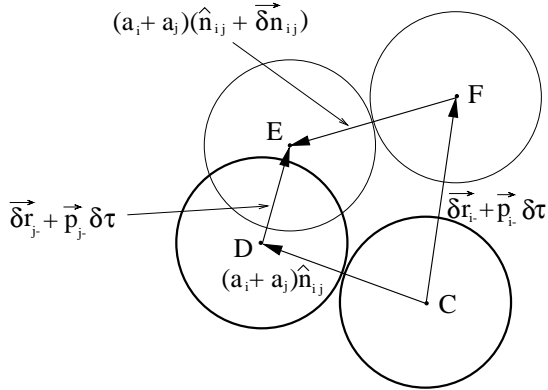}
\end{center}
\caption{Collision between the $i$-th and the $j$-th sphere.
Thick-lined spheres are on the reference trajectory whereas the
thin-lined spheres are on the adjacent trajectory.\label{fig4}}
\end{figure}

\begin{figure}
\begin{center}
\includegraphics[width=0.4\textwidth]{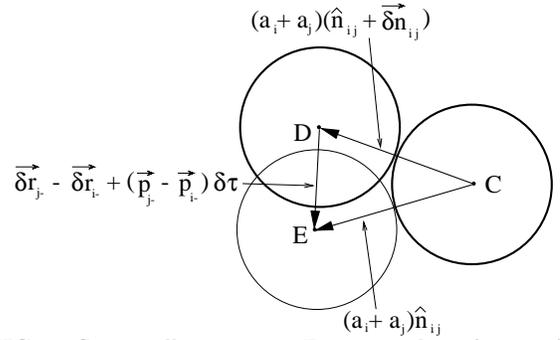} \nobreak
\caption{Same collisions as in Fig.~\ref{fig4}, in the reference frame
of the $i$-th sphere.\label{fig5}}
\end{center}
\end{figure}

\noindent
and $\dot{\vec{\Gamma}}_{\pm}$ describes the equations of motion,
Eq.~(\ref{e2.7}), right after(before) the collision at A, i.e.,
\begin{eqnarray}
   \dot{\vec{\Gamma}}_{\pm}= \left[\begin{array}{c}
	{\dot{\vec{R}}_{\pm}}\\ {{\dot{\vec{V}}_{\pm}}} \end{array}
	\right] = \left[\begin{array}{c} {\vec{V}_{\pm}}\\
	{\alpha_{0}\gamma{\mathbf
	C}\vec{R}_{\pm}-\alpha_{0}\vec{V}_{\pm}} \end{array} \right].
\label{ea4.16}
\end{eqnarray}
Having calculated the quantity $\ds{\frac{\partial{\mathbf
Q}}{\partial\vec{\Gamma}_{-}}}$ from Eq.~(\ref{ea4.13}),  \bleq
\renewcommand{\theequation}{A\arabic{equation}}
\begin{eqnarray}
   \frac{\partial{\mathbf Q}}{\partial\vec{\Gamma}_{-}}=
	\left[\begin{array}{cc} {{\mathbf I}\,\,\,\,\,\,\,\,\,}
	&{{\mathbf 0}}\\ \!\!{-\,2\,\vec{V}_{-}\!\cdot\!\left[
	\displaystyle{{{\frac{\partial\hat{N}_{ij}}
	{\partial\vec{R}_{-}}}}\,\hat{N}_{ij}\,+\,\hat{N}_{ij}
	{{\frac{\partial\hat{N}_{ij}}{\partial\vec{R}_{-}}}}}
	\right]\,\,\,\,\,\,\,\,\,} &{{\mathbf
	I}-2\hat{N}_{ij}\hat{N}_{ij}}\!  \end{array}\right]\,,
\label{ea4.17}
\end{eqnarray}
where each entry of the matrix on the r.h.s. of Eq.~(\ref{ea4.17}) is
a $3N\!\times\!3N$ matrix, the expressions for $\vec{\delta R}_{+}$
and $\vec{\delta V}_{+}$ can be obtained from Eq.~(\ref{ea4.14}) as
\begin{eqnarray}
   \vec{\delta R}_{+}= \vec{\delta R}_{-}\, -\,2\,(\vec{\delta
	R}_{-}\cdot\hat{N}_{ij})\,\hat{N}_{ij}\,,
\label{ea4.18}
\end{eqnarray}
and
\begin{eqnarray}
   \vec{\delta V}_{+}= {\mathbf A}\cdot\vec{\delta R}^{*}_{-}
	+({\mathbf I}-2\hat{N}_{ij}\hat{N}_{ij})\vec{\delta V}_{-}
	+({\mathbf I}-2\hat{N}_{ij}\hat{N}_{ij})
	[\alpha_{0}\gamma{\mathbf C}\vec{R}_{-}
	-\alpha_{0}\vec{V}_{-}]\delta\tau -[\alpha_{0}\gamma{\mathbf
	C}\vec{R}_{+}-\alpha_{0}\vec{V}_{+}]\delta\tau,
\label{ea4.19}
\end{eqnarray}
where ${\mathbf A}=-2\vec{V}_{-}
\cdot\bigg[{\ds{\frac{\partial\hat{N}_{ij}}
{\partial\vec{R}_{-}}}}\,\hat{N}_{ij}\,+\,\hat{N}_{ij}
{\ds{\frac{\partial\hat{N}_{ij}}{\partial\vec{R}_{-}}}}\bigg]$ and
$\vec{\delta R}^{*}_{-}\, =\,(\vec{\delta R}_{-}\,
+\,\vec{V}_{-}\,\delta\tau)$. At this point, we use Eq.~(\ref{e2.9})
and obtain
\begin{eqnarray}
   ({\mathbf{I}}-2\hat{N}_{ij}\hat{N}_{ij})\,
	[\,\alpha_{0}\gamma{\mathbf C}\vec{R}_{-}\,
	-\,\alpha_{0}\vec{V}_{-}\,]\,\delta\tau\,
	-\,[\,\alpha_{0}\gamma{\mathbf C}\vec{R}_{+}\,
	-\,\alpha_{0}\vec{V}_{+}\,]\,\delta\tau =
	-\,2\alpha_{0}\gamma\,(\hat{N}_{ij}\cdot{\mathbf
	C}\vec{R}_{-})\, \hat{N}_{ij}\,\delta\tau\,.
\label{ea4.20}
\end{eqnarray} 
Following Appendix B of \cite{Panja_preprint}, the term ${\mathbf
A}\cdot\vec{\delta R}^{*}_{-}$ can be expressed as
\begin{equation}
   {\mathbf A}\cdot\vec{\delta R}^{*}_{-}=
-2\,[(\vec{V}_{-}\cdot\vec{\delta N}_{ij})\hat{N}_{ij}
+(\vec{V}_{-}\cdot\hat{N}_{ij})\vec{\delta N}_{ij}]\,,
\label{ea1}
\end{equation}
\eleq

\noindent
where
\begin{eqnarray}
   \vec{\delta N}_{ij}=
	\frac{1}{\sqrt{2}}\,(\vec{0},\,\vec{0},\,.\,.,\, \vec{\delta
	n}_{ij},\ldots-\vec{\delta n}_{ij},\ldots \vec{0}\,)\,,
\label{ea2}
\end{eqnarray}
satisfying $\hat{n}_{ij}\cdot\vec{\delta n}_{ij}=0$. This
orthogonality condition between $\hat{n}_{ij}$ and $\vec{\delta
n}_{ij}$ also implies that $\hat{N}_{ij}\cdot\vec{\delta
N}_{ij}=\,0$. To obtain an expression for $\vec{\delta n}_{ij}$, we
need to take a look at Figs.~\ref{fig4} and \ref{fig5}. Figure
\ref{fig4} describes, in the laboratory frame, the binary collision
process between the $i$-th and the $j$-th sphere on the reference and
adjacent trajectories; the thick-lined spheres are on the reference
trajectory whereas the thin-lined spheres are on the adjacent
trajectory.  Figure \ref{fig5} describes the same binary collision
process in the reference frame of the $i$-th sphere (with center
C). In Fig.~\ref{fig4}, the thick-lined $j$-th sphere (with center D)
on the left depicts the collision situation on the reference
trajectory and the thin-lined $j$-th sphere (with center E) on the
left depicts the collision situation on the adjacent trajectory.
\vspace{1em}
\noindent Clearly, in Fig.~\ref{fig5}, the infinitesimal vector
$\overrightarrow{\mbox{DE}}$ is given by
\begin{eqnarray}
   \vec{\delta r}^{*}_{ij}= \vec{\delta r}_{j-}-\vec{\delta r}_{i-}\,
	+\,(\vec{v}_{j-}-\vec{v}_{i-})\,\delta\tau
\label{ea3}
\end{eqnarray}
and since the lengths of both the lines CD and CE are $a_{i}+a_{j}$
($a_{i}$ and $a_{j}$ are the radii of the $i$-th and the $j$-th sphere
respectively), we have
\begin{eqnarray}
   \vec{\delta n}_{ij}= \frac{1}{a_{i}\,+\,a_{j}}\,\vec{\delta
	r}^{*}_{ij}\,.
\label{ea4}
\end{eqnarray}

Let us define a $3N\!\times\!3N$ matrix ${\mathbf U}$ composed of
$N\times N$ blocks of $3\!\times\!3$ matrices, such that, in terms of
the block indices the only non-zero entries of ${\mathbf U}$ are
\begin{eqnarray}
   {\mathbf U}_{ii}= -{\mathbf U}_{ij} = -{\mathbf U}_{ji} = {\mathbf
	U}_{jj} = -{\mathbf I}\,,
\label{ea5}
\end{eqnarray}
where ${\mathbf I}$ is the $3\!\times\!3$ unit matrix. One can now
write, using Eqs.~(\ref{ea2}-\ref{ea5}) and Eq.~(\ref{ea4.12}) that
\bleq
\begin{eqnarray}
   \vec{\delta N}_{ij}= \frac{1}{\sqrt{2}(a_{i}\,+\,a_{j})}[\,{\mathbf
	U}\cdot\vec{\delta R}_{-} +\,{\mathbf
	U}\cdot\dot{\vec{\Gamma}}_{-}\,\delta\tau] =
	\frac{1}{\sqrt{2}(a_{i}\,+\,a_{j})}\left[\, {\mathbf
	U}-\frac{({\mathbf U}\cdot\vec{V}_{-})\hat{N}_{ij}}
	{\vec{V}_{-}\cdot\hat N_{ij}}\,\right]\cdot\vec{\delta
	R}_{-}\,.
\label{ea6}
\end{eqnarray}
{}From Eqs.~(\ref{ea1}) and Eq.~(\ref{ea6}), we finally have
\begin{eqnarray}
   {\mathbf A}\cdot\vec{\delta R}^{*}_{-}&=&
	\frac{\sqrt{2}}{a_{i}\,+\,a_{j}}\Bigg[ ({\mathbf
	U}\cdot\vec{V}_{-})\hat{N}_{ij}\,  +\,\hat{N}_{ij}({\mathbf
	U}\cdot\vec{V}_{-})\,
	-\,(\hat{N}_{ij}\cdot\vec{V}_{-}){\mathbf U}\,
	+\,\frac{\vec{V}_{-}\cdot{\mathbf U}\cdot\vec{V}_{-}} {\vec
	V_-\cdot\hat N_{ij}}\hat N_{ij}\hat N_{ij}
	\Bigg]\cdot\vec{\delta R}_{-} \nonumber \\ &=& (\,{\mathbf
	I}-2\hat{N}_{ij}\hat{N}_{ij}\,)\, {\mathbf W}\cdot\vec{\delta
	R}_{-}\,.
\label{ea7}
\end{eqnarray}
\eleq
\noindent
where ${\mathbf W}$ is a $3N\!\times\!3N$ symmetric matrix.  Finally,
using Eqs.~(\ref{ea4.12}), (\ref{ea4.18}-\ref{ea4.20}) and
(\ref{ea7}), the  expression for ${\mathbf M}_{ij}$ can be obtained as
\begin{eqnarray} 
   {\mathbf M}_{ij}= ({\mathbf
	I}-2\hat{N}_{ij}\hat{N}_{ij})\,\left[\begin{array}{cc}
	{{\mathbf I}\,\,\,\,\,\,\,\,}&{\mathbf 0}\\ {\mathbf
	R\,\,\,\,\,\,\,}&{\mathbf I}\!  \end{array}\right]\,,
\label{ea4.21}
\end{eqnarray}
where
\begin{equation}
{\mathbf R}=\!\dst{{\mathbf W}-\!2\alpha_{0}\gamma\,
\dst{\frac{\hat{N}_{ij}\cdot{\mathbf C}\vec{R}_{-}}
{\vec{V}_{-}\cdot\hat{N}_{ij}}}\,\hat{N}_{ij}\hat{N}_{ij}} \,.
\label{eaR}
\end{equation}

\ecols


\begin{thebibliography}{00}
\bibitem{Noteonthermodynamiclimit} In reality, turbulence develops if
the system's dimension in the direction on the gradient of velocity
becomes too large, but the iso-kinetic and the constant $\alpha$
thermostats stabilize the linear flow field, even for
$N\rightarrow\infty$.
\bibitem{sufficient} This condition has been quoted as a necessary
condition for an exact CPR in Ref.~\cite{SEI_chaos_98}, but it is
actually a sufficient condition.
\end{thebibliography}
\end{document}